\documentclass[aps,pre,twocolumn,showpacs,floatfix]{revtex4}
\usepackage{epsfig}
\usepackage{psfrag}
\begin{document}
\title{Break-up of shells under explosion and impact}

\author{ F.\ K.\ Wittel${}^1$, F.\ Kun${}^{2}\footnote{Electronic 
address:feri@dtp.atomki.hu}$, H.\ J.\ Herrmann${}^3$, and B.\ H.\
Kr\"oplin${}^1$} 
\affiliation{
${}^1$Institute of Statics and Dynamics of Aerospace Structures,
University of Stuttgart, Pfaffenwaldring 27, 70569 Stuttgart,
Germany\\  
${}^2$Department of Theoretical Physics,
University of Debrecen, P.\ O.\ Box:5, H-4010 Debrecen, Hungary \\
${}^3$ICP, University of Stuttgart, Pfaffenwaldring 27, D-70569
Stuttgart, Germany} 
\date{\today}
   
\begin{abstract}
A theoretical and experimental study of the fragmentation
of closed thin shells made of a disordered brittle material is
presented.   
Experiments were performed on brown and white hen egg-shells
under two different loading conditions: fragmentation due to an
impact with a hard wall and explosion by a combustion mixture
giving rise to power law fragment size distributions.
For the theoretical investigations a three-dimensional discrete
element model of shells is constructed. Molecular dynamics simulations
of the two loading cases resulted in power law fragment mass
distributions in satisfactory 
agreement with experiments. Based on large scale simulations we give
evidence that power law distributions arise due to an underlying
phase transition which proved to be abrupt and continuous for
explosion and impact, respectively. Our results demonstrate 
that the fragmentation of closed shells defines a universality class
different from that of two- and three-dimensional bulk systems. 
\end{abstract}
 
\pacs{64.60.-i, 64.60.Ak, 46.30.Nz}

\maketitle

Closed shells made of solid materials are often used in every day
life, industrial applications and engineering practice as containers,
pressure vessels or combustion chambers. From a structural point of view
aircraft vehicles, launch vehicles like rockets and building blocks of
a space station are also shell-like systems, and even
certain types of modern buildings can be considered as shells. The
egg-shell as nature's oldest container proved to be a reliable
construction for protecting life. 
In most of the applications shell-like constructions operate under an
internal pressure much higher than the surrounding one. 
Hence, careful design and optimization of structural and
material properties is required to ensure the stability and
reliability of the system. Closed shells usually fail due to an excess
internal load 
which can arise either as a result of slowly driving the system above its
stability limit during its usage or service time, or by a pressure
pulse caused by an explosive shock inside the shell. Due to the
widespread applications, the failure of shell systems is a very
important scientific and technological problem which has also 
an enormous social impact due to the human costs arising, for
instance, in accidental events.

Fragmentation, {\it i.e.} the breaking of particulate materials into
smaller pieces is abundant in nature and underlies several industrial
processes, which attracted a continuous interest 
in scientific and engineering research over the past decades \cite{frag_book,gilvary,ice,exp_tur,explode,exp_colli1,exp_colli2,exp_colli3,exp_bohr,exp_meibom,exp_inao,exp_kadono,kadono,katsuragi,katsuragi1,porose}.
Fragmentation phenomena can be observed on a broad range of
length scales ranging from the collisional evolution 
of asteroids and meteor impacts on the
astrophysical scale, through geological phenomena and industrial
applications on the intermediate scale down to the break-up of large
molecules and heavy nuclei on the  
atomic scale. In laboratory experiments on the fragmentation of solids,
the energy input is usually achieved by shooting a projectile into a solid
block
\cite{gilvary,exp_colli2,exp_colli3,exp_inao,exp_kadono,kadono},
making an explosion inside the sample \cite{exp_tur,explode} or by the collision 
of macroscopic bodies (free fall impact)
\cite{ice,exp_colli1,exp_bohr,exp_meibom,katsuragi,katsuragi1}. Due to
the violent nature of  
the process, observations on fragmenting systems are often restricted
to the final state, making the fragment size (volume, mass, charge,
...) to be the main characteristic quantity. 
The most striking observation on fragmentation is that the
distribution of fragment sizes shows a power 
law behavior, independently on the way of imparting energy, relevant
microscopic interactions and length scales involved, with an exponent
depending only on the dimensionality of the system \cite{exp_tur,gilvary,explode,exp_colli1,exp_colli2,exp_colli3,exp_bohr,exp_meibom,exp_inao,exp_kadono,kadono,katsuragi,katsuragi1,campi3,ice,porose}. During the past
years experimental
\cite{exp_tur,gilvary,explode,exp_colli1,exp_colli2,exp_colli3,exp_bohr,exp_meibom,exp_inao,exp_kadono,kadono,katsuragi,katsuragi1,campi3,ice,porose} and
theoretical \cite{automata,soc,zhang,astrom2,astrom3,botet,engelman1,gonzalo,holian1,astrom1,ching,diehl,bershadski,bershadski1,bershadski2,shatter1,kun1,bhupal,ferenc1,ferenc2,falk,thornton,benz,potapov2} efforts focused on the 
validity region and the reason of the observed universality in 1, 2,
and 3 dimensions. Detailed studies  
revealed that universality prevails for large enough input energies
when the system falls apart into small enough pieces
\cite{exp_bohr,exp_meibom,exp_inao,exp_kadono,kadono,katsuragi,katsuragi1,porose,automata,soc,zhang,astrom2,astrom3,botet,engelman1,gonzalo,holian1,astrom1}, however, at
lower energies a systematic dependence of the exponent on the input
energy was evidenced \cite{ching,diehl}. Recent investigations on the
low energy limit of fragmentation suggest that the power law
distribution of fragment sizes arises due to an underlying critical
point \cite{astrom2,astrom3,astrom1,kun1,bhupal,falk}. 

Besides the industrial and social impact of the failure of shell like
systems, they are also of high scientific importance for the understanding
of fragmentation phenomena. Former studies on fragmentation have
focused on the behavior of bulk systems in one, two and three
dimensions under impact and explosive loading, however, hardly
any studies have been devoted to fragmentation of shells
\cite{falk}. The 
peculiarity of the break-up of closed shells originates from the
fact that the local structure is inherently two-dimensional, however,
the dynamics of the systems, the motion of material elements,
deformation and stress states are three-dimensional which allows for a
rich variety of failure modes \cite{falk}.
 
In this paper we present a detailed experimental and theoretical
study of the fragmentation of closed solid shells arising due to an
excess load inside the shell.
Experiments were performed on brown and white hen egg-shells
under two different loading conditions: fragmentation due to an
impact with a hard wall and explosion by a combustion mixture have
been considered resulting in power law fragment size distributions.
For simplicity, our theoretical study is restricted to spherical
shells such that a three dimensional discrete element model of
spherical shell systems was worked out.  
In molecular dynamics simulations of the two loading cases,
power law fragment mass distributions were obtained in satisfactory 
agreement with experiments. Based on large scale simulations we give
evidence that power law distributions arise due to an underlying
phase transition which proved to be abrupt for explosion and continuous
for impact. Analyzing the energetics
of the explosion process in the two loading 
cases and the evolution of the fragment mass distributions we
demonstrate that the fragmentation of closed shells defines a
new universality class different from that of two- and three-dimensional
bulk systems.  

\section{Experiments}
Hen eggs provide an excellent possibility for the study of
fragmentation of thin brittle shells of disordered materials 
with the additional advantages of being cheap and easy to handle,
making the patience of scientists the only limiting factor for the
subsequent improvement of the experimental results.  
Our experiments were performed on ordinary brown and white
egg-shells. 
In the preparations, first two holes of
regular circular shape were drilled on the bottom and top of the egg
through which the content of the egg was blown-out. The inside
was carefully washed and rinsed out several times and finally the empty
shells were dried in a microwave oven to get rid of all moisture of the
egg-shell. The drying process proved to be essential to ensure that
the cuticula, which cannot be blown out, competely looses its
toughness. 

\begin{figure}
\begin{center}
\epsfig{bbllx=20,bblly=20,bburx=580,bbury=250,file=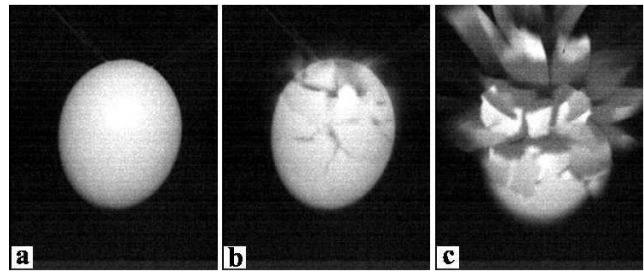,
  width=8.5cm}
 \caption{\small Time evolution of the explosion of an egg-shell,
   consecutive snapshots taken by a high speed camera. The time
   difference between the snapshots is 0.001 sec. 
}
\label{fig:eggsplode}
\end{center}
\end{figure}
In the impact experiments intact egg-shells are catapulted onto the
ground at a high speed using a simple setup of rubber bands. 
The experimental setup provided a relatively high energy
impact without the possibility of varying the imparted energy. 
The eggs are shot directly into a plastic bag touching the
ground so that no fragments are lost for further evaluation. 

In the explosion experiment initially the egg-shell is flooded with
hydrogen and 
hung vertically inside a plastic bag. The combustion reaction is
initiated by igniting the 
escaping hydrogen on the top of the egg. The
hydrogen immediately reacts with the Oxygen which is also
drawn up into the egg through the bottom hole, mixing with the
remaining hydrogen. When enough air has entered to form a combustible
mixture inside the egg, the flame back-fires through the top hole and
starts the very quick exothermic reaction. The experiment is
carried out inside a soft plastic bag so that secondary fragmentations due to
fragment-wall collisions do not occur, furthermore, no
pieces were lost after explosion.
Since the pressure which builds up during combustion
can slightly be changed by the hole size, {\it i.e} the smaller the
hole, the higher the pressure at the explosion,
we performed several series of experiments with hole diameters $d$ between
1.2 and 2.5 millimeter. The limit values have practical reasons:
a drilling nail of large diameter typically breaks the eggs-shell,
on the other hand it is extremely difficult to blow out an egg through
a hole of diameter 1 mm or less.

\begin{figure}
\begin{center}
\epsfig{bbllx=20,bblly=20,bburx=575,bbury=375,file=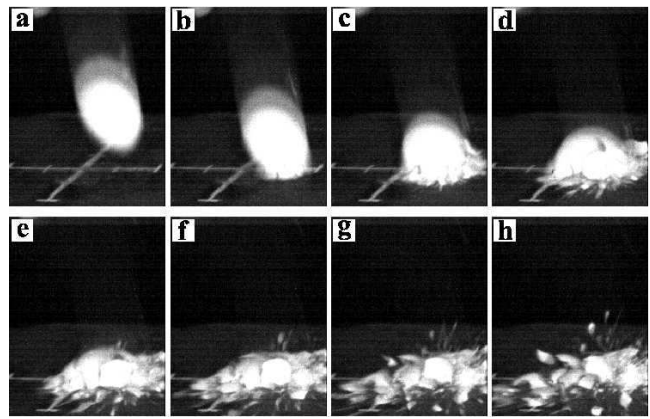,
  width=8.5cm}
 \caption{\small Time series of the impact of an egg-shell with the hard
   ground. The consecutive snapshots were taken by a high speed camera
   of 1 kHz. 
}
\label{fig:impact}
\end{center}
\end{figure}
It is possible to follow the time evolution of the explosion and
impact processes by means of a high speed camera under well controlled
conditions. Three consecutive snapshots  of the explosion
process are presented in Fig.\ \ref{fig:eggsplode} taken by a camera of
1000 Hz frequency.
The ignition took place at the top of the egg in Fig.\
\ref{fig:eggsplode}$a)$. The instant of back-firing and 
the initiation of combustion is captured in Fig.\
\ref{fig:eggsplode}$b)$, while in Fig.\ \ref{fig:eggsplode}$c)$
already the flying pieces can be seen. Based on the
snapshots the total duration of an explosion is estimated to be of the
order of 1 millisecond. 

In the impact experiment the egg hits the
ground in the direction of its longer axis, as it is illustrated by
the picture series of Fig.\ \ref{fig:impact}. After hitting
the ground (Fig.\ \ref{fig:impact}$b$), the egg suffers gradual collapse
as it moves forward (Fig.\ \ref{fig:impact}$c-h$) making the impact
process relatively longer compared to the explosion. 

\begin{figure}
\begin{center}
\epsfig{bbllx=30,bblly=20,bburx=585,bbury=500,file=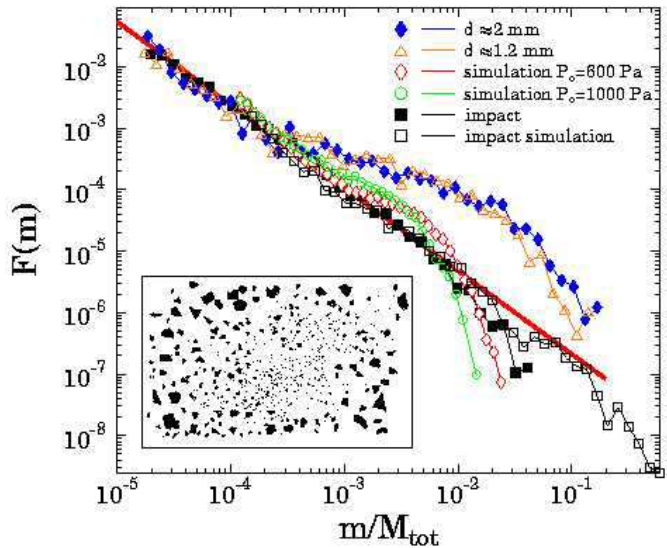,
    width=9.0cm}
 \caption{\small Comparison of fragment mass distributions obtained by
   explosion experiments with two hole sizes and the impact
   experiment to the simulation results. The inset shows a typical
   scanned set of fragments. 
}
\label{fig:exp_imp}
\end{center}
\end{figure}

The resulted egg-shell pieces are then carefully collected and placed
on the tray of a scanner without overlap. In the scanned image
fragments are seen as black spots on a white background
and were further analyzed by a cluster searching code. 
In the inset of Fig.\ \ref{fig:exp_imp} an example of scanned pieces
of an  impact experiment is shown where the
broad variation of sizes can also be noticed with the naked eye.
A dusty phase of shattered pieces \cite{redner} was also observed in the experiments
with fragment 
sizes falling in the order of the pixel size of the scanner. 
The mass $m$ of fragments was determined as the number of pixels
in the scanned image. Since shattered fragments were also
comparable to normal dust pieces 
in the air, they were excluded in the analysis by setting the lower
cut-off of fragment masses to a few pixels. 

As the main quantitative result of the experiments we evaluated the
mass distribution of fragments $F(m)$ which is defined so that
$F(m)\Delta m$ provides the probability of finding a fragment 
with mass falling between $m$ and $m+\Delta m$. 
Fig.\ \ref{fig:exp_imp}
presents the fragment mass distributions $F(m)$ for impact
and explosion experiments averaged over 10-20 egg-shells for each
curve. For the impact experiment, a power law behavior of
the fragment mass distribution 
\begin{eqnarray}
 F(m) \sim m^{-\tau}
\end{eqnarray}
can be observed over three orders of magnitude where
the value of the exponent can be determined with high precision
to $\tau = 1.35 \pm 0.02$. Explosion
experiments result also in a power law distribution of the same value
of $\tau$ for small fragments with a
relatively broad cut-off for the large ones.
Smaller hole diameter $d$ in Fig.\ \ref{fig:exp_imp}, {\it i.e.}
higher pressure, gives rise to a
larger number of fragments with a smaller cut-off mass and a faster
decay of the distribution $F(m)$ at the large fragments. Comparing the
number of fragments obtained, the ratio of the pressure values
in the explosions at hole diameters $d=1.2$ and 2.0 mm, presented
in Fig.\ \ref{fig:exp_imp}, was estimated to be about 1.6.
Note that the relatively
small value of the exponent $\tau$ can indicate a cleavage mechanism
of shell fragmentation and is significantly different from the
experimental and theoretical results on fragmenting two-dimensional
bulk systems where $1.5 \leq \tau \leq 2$ has been found
\cite{exp_tur,exp_inao,exp_kadono,exp_bohr,exp_meibom,katsuragi,katsuragi1,astrom2,astrom3,kun1,bhupal,ferenc1,ferenc2},
and from the three-dimensional ones where $\tau > 2$ is obtained \cite{exp_tur,exp_colli2,exp_colli3,thornton,benz}.

\section{Simulations}
Most of the theoretical studies on fragmentation relay on
large scale computer simulations since capabilities of analytic
approaches are rather limited in this field due to the
complexity of the break-up process. Over the past years the
Discrete Element Method (DEM) proved to be a very efficient numerical
technique for fragmentation phenomena \cite{astrom2,astrom3,astrom1,kun1,bhupal,ferenc1,ferenc2,falk,thornton,benz,potapov2} since it has the ability to
handle large deformations arising dynamically, and naturally captures
the propagation and interaction of a large number of simultaneously
growing cracks, which is 
essential for fragmentation. 

\begin{figure}
\begin{center}
\epsfig{bbllx=20,bblly=20,bburx=570,bbury=690,file=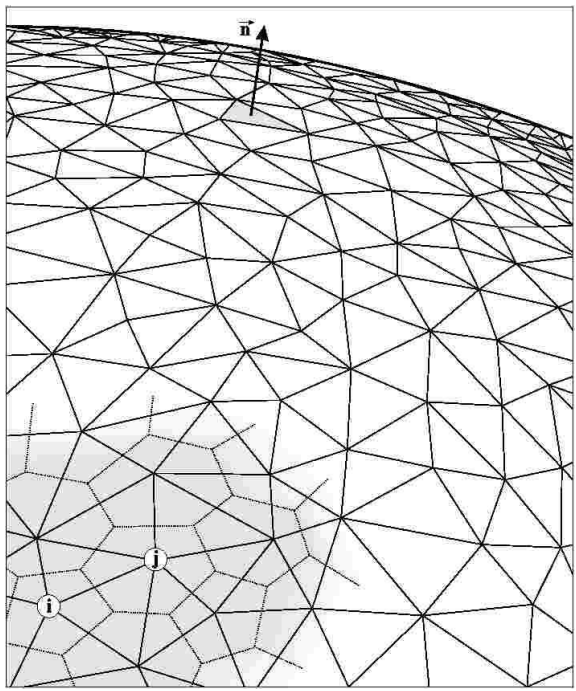,
  width=7.5cm}
 \caption{\small Example of the Delaunay triangulation of the
   spherical surface. The dual Voronoi lattice is also shown in the
   lower left quadrant. 
}
\label{fig:mesh}
\end{center}
\end{figure}

In order to investigate the fragmentation
of spherical shells we constructed a three-dimensional discrete
element model such that the surface of the unit sphere is discretized
into randomly shaped triangles (Delaunay triangulation) by throwing
points randomly and independently on the surface
\cite{moukarzel,lauritsen}. 
Based on the triangulation, the dual Voronoi tessellation of the
surface is also carried 
out as is illustrated in Fig.\ \ref{fig:mesh}. The nodes of the
triangulation represent point-like material  
elements in the model whose mass is defined by the area of the Voronoi
polygon assigned to it \cite{ferenc1,moukarzel,lauritsen}. The bonds between
nodes are assumed to be 
springs having linear elastic behavior up to failure. 
Disorder is introduced in the model solely by the randomness of the
tessellation so that the mass assigned to the nodes, the
length and cross-section of the springs are determined by the
tessellation (quenched structural disorder). After prescribing the
initial conditions of a specific fragmentation process, the
time evolution of the system is 
followed by solving the equation of motion of nodes by a
Predictor-Corrector method of fourth order
\begin{eqnarray}
\label{eq:eom}
m_i \ddot{\vec{r}}_i = \vec{F}_i^s + \vec{F}_i^{ext} + \vec{F}_i^{d},
\ \ \ i=1, \ldots N, 
\end{eqnarray}
where $\vec{F}_i^s$ is the sum of forces exerted by the springs
connected to node $i$, and $\vec{F}_i^{ext}$ denotes the external driving
force, which depends on the loading condition. To
facilitate the relaxation of the system at the end of the
fragmentation process, a small viscous damping force $\vec{F}_i^{d}$
was also introduced in Eq.\ (\ref{eq:eom}).

In order to account for crack formation in the model springs are
assumed to break when their deformation $\varepsilon$ exceeds a
certain breaking 
threshold $\varepsilon_c$. A fixed threshold value $\varepsilon_c =
0.03$ is set for all the springs resulting in a random sequence of
breakings due to the disordered spring properties. The breaking criterion
is evaluated at each iteration step and those springs which fulfill
the condition are removed from the simulation. As a result of
successive spring breakings cracks nucleate, grow and merge on the
spherical surface which can give rise to a complete break-up of the
shell into smaller pieces. 

Fragments of the shell are defined in the
model as sets of nodes (material elements) connected by the remaining
intact springs.   
The process is stopped when the system has attained a relaxed state,
{\it i.e.} when 
there is no spring breaking over a large number of iteration
steps. The main advantage of DEM is that it makes it
possible to monitor a large number of microscopic physical quantities
during the course of the simulation which are hardly accessible
experimentally, providing a deep insight into the
fragmentation process. With the present computer capacities, DEM models
can be designed to be realistic so that the simulation results can
even complement the experimental 
information extending our understanding. The most important parameter
values used in our simulations are summarized in Table \ref{tab:table1}.

In computer simulations two different
ways of loading have been considered which model the experimental
conditions and represent limiting cases of energy input rates:
(i) {\it pressure pulse} and (ii) {\it impact} load. A pressure pulse
in a shell is carried out by imposing a fixed internal pressure $P_o$
from which the forces $\vec{F}_j^{ext}$ acting on the triangular
surface elements are calculated as 
\begin{eqnarray}
\label{eq:forces}
\vec{F}_j^{ext} = P_o A_j \vec{n}_j,
\end{eqnarray}
where $A_j$ denotes the actual area of triangle $j$ and the force
points in the direction of the local normal $\vec{n}_j$, see also Fig.\
\ref{fig:mesh}. The force $F_j^{ext}$ is 
equally shared by the three nodes of the triangle for which the
equation of motion Eq.\ (\ref{eq:eom}) is solved. 
Since the surface area of the shell increases, the expansion under constant
pressure implies a continuous increase of the driving force and of the
imparted energy.

The impact loading realizes the limiting case of instantaneous energy
input by giving a fixed initial radially oriented velocity $v_o$ to
the material elements and
following the resulted time evolution of the system by solving the
equation of motion Eq.\ (\ref{eq:eom}). 
The control parameter of the system which determines the final outcome
of the process are the fixed pressure $P_o$ and the initial kinetic
energy $E_o$ for the pressure pulse and impact loading, respectively. 
\begin{table}
\begin{center}

\begin{tabular}{|c|c|c|c|}

\hline
$Parameter$           & $Symbol$    & $Unit$   & $Value$  \\
\hline
Initial radius        &   $R$       &  $m$     & 1    \\
Initial volume        &   $V_o$     &  $m^3$   & 4.19 \\
Initial Surface       &   $A_o$     &  $m^2$   & 12.56 \\
Shell thickness       &   $th$      &  $m$     & $5\cdot10^{-5}$ \\
Total mass            &   $M_{tot}$ &  $kg$    & 0.816 \\
Num.\ of triangles    &   $N_t$     &          & $\approx 44000$ \\
Num.\ of nodes        &   $N_n$     &          & $\approx 21000$ \\
\hline 
\hline
Mass density          &   $\rho$    &  $kg/m^3$& $1300$ \\       
Time step             &   $\Delta t$&  $s$     & $3\cdot10^{-7}$ \\
Damping coeff.        &   $\gamma_d$&  $kg/s$  & $0.1$   \\
Spring Young modul.\  &   $Y$       &  $N/m^2$ & $10^9$  \\
\hline
\end{tabular}
\caption{ \label{tab:table1} Parameter values used in the simulations.
}
\end{center}
\end{table}

\section{The break-up process}
In the simulations, in both loading cases the spherical shell 
is initially completely
stress free with no energy stored in deformation. When a
constant pressure is imposed the total
energy $E_{tot}$ of the shell increases due to the work done 
by the filling gas
\begin{eqnarray}
 E_{tot}(V) = \int_{V_o}^{V} P_o dV = P_o \Delta V,
\end{eqnarray}
where $V$ denotes the actual volume during the expansion and $\Delta
V$ is the volume change with respect to the initial state $V_o$. 
\begin{figure}
\begin{center}
\epsfig{bbllx=20,bblly=20,bburx=570,bbury=510,file=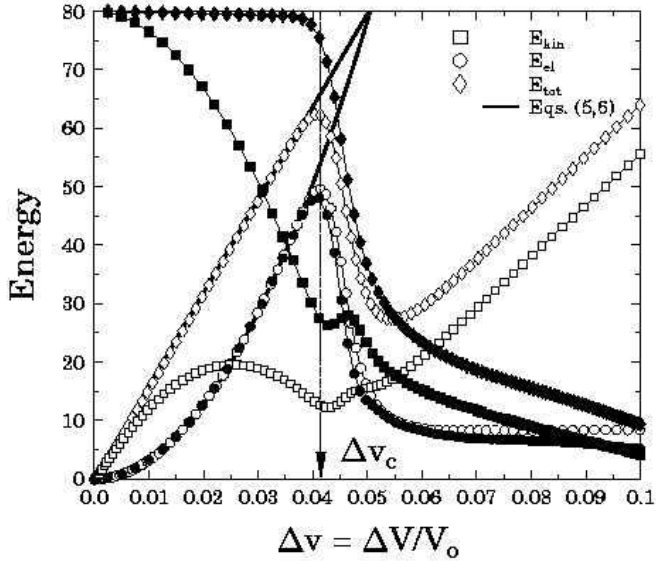,
  width=9.1cm}
 \caption{\small The kinetic $E_{kin}$, elastic $E_{el}$, and total
   $E_{tot}$ energies as a function of the relative volume change
   $\Delta v$. Open symbols stand for expansion under constant
   pressure while the filled ones characterize the impact loading. The
   sudden drop of the total and elastic energy at $\Delta v_c$
   indicates the rapid break-up of the system. For $E_{tot}$ of the
   pressure loading the thick solid line follows Eq.\ (\ref{eq:etot}),
   while for $E_{el}$ the function given by Eq.\ (\ref{eq:eel}) was
   fitted with $C=312000$ as a parameter.
}
\label{fig:energy_both}
\end{center}
\end{figure}
The total energy can be written as the sum of the kinetic energy of
material elements $E_{kin}$ and of the elastic energy $E_{el}$ stored
in deformation, $E_{tot} = E_{kin} + E_{el}$, where $E_{el}$ is
proportional to the change $\Delta A$ of surface area $A$ of the
expanding sphere with respect to the initial surface $A_o$. 
Introducing the relative volume change $\Delta v =
\frac{\Delta V}{V_o}$ as an
independent variable, the total energy and the elastic energy
can be cast in the form
\begin{eqnarray}
 \label{eq:etot}
 E_{tot} &=& P_oV_o\Delta v, \\
 E_{el}  &=& C\left[\left(\Delta v+1\right)^{1/3}-1\right]^2,
 \label{eq:eel} 
\end{eqnarray}
where the surface change $\Delta A$ was expressed in terms of $\Delta
v$. Furthermore, the parameter $C$ of the system depends on the
properties of 
the triangulation and the characteristic physical quantities of
springs (Young modulus, length, thickness). 
\begin{figure}
\begin{center}
\epsfig{bbllx=20,bblly=20,bburx=580,bbury=770,file=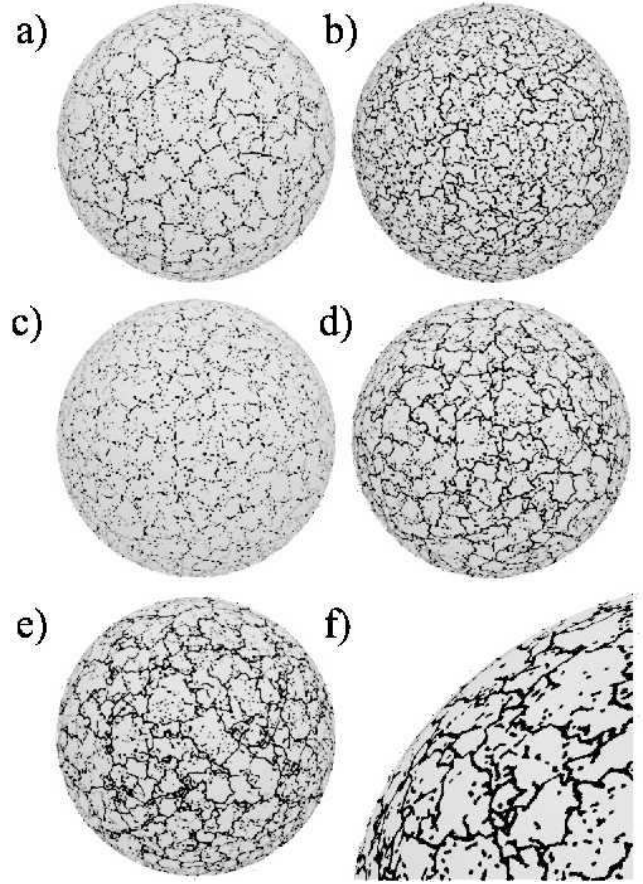,
  width=8.8cm}
 \caption{\small 
Cracks on the shell surface. 
Final states of impact experiments at energies $E_o/E_c \approx 0.8$ $a)$ and
2.8 $b)$. Time evolution of the cracking
process under a constant pressure of $P_o/P_c \approx 4.0$ $c,d)$ until the final
relaxed state is reached $e)$, with a magnified view of fragments $f)$. 
Particle positions are projected back to their
initial state on the surface. Fragments are identified as shell
pieces surrounded by cracks. 
}
\label{fig:cracks}
\end{center}
\end{figure}
It is interesting to note
that there exists a specific 
pressure value $P_o^*$ below which the expansion always stops at a
maximum volume change $\Delta v_{max}$ depending on $P_o$, however,
for $P_o > P_o^*$ the 
expansion keeps always accelerating. For a
given $P_o < P_o^*$ the value of $\Delta v_{max}$ can be determined
from the condition $E_{tot} = E_{el}$ so that 
\begin{eqnarray}
\label{eq:p_star}
 P_oV_o\Delta v_{max} = C\left[\left(\Delta v_{max}+1 \right)^{1/3}-1
 \right]^2,
\end{eqnarray}
and $P_o^*$ can be identified as the highest pressure for which Eq.\
(\ref{eq:p_star}) 
can be solved for $\Delta v_{max}$. Usually $\Delta v_{max}$ can only
be realized at low pressure values, because at higher pressures the
system suffers complete break-up much below  $\Delta v_{max}$, due to
the finite strength of the springs. 
Fig. \ref{fig:energy_both} illustrates the evolution of the total
$E_{tot}$, kinetic $E_{kin}$, and elastic $E_{el}$ energies as a
function of $\Delta v$ for both pressure and impact loading.
In the case of pressure loading it can be observed that the total
energy $E_{tot}$ extracted from the 
simulations agrees well with the analytic prediction of Eq.\
(\ref{eq:etot}).
The numerical value of the multiplication factor $C$ of the
elastic energy was obtained by fitting the expression Eq.\
(\ref{eq:eel}) to the curve of $E_{el}(\Delta v)$ in the
figure. Due to the constant pressure, the total force $F$ acting on
the shell is proportional to the actual surface 
area $F \sim A\cdot P_o$ so that the system is driven by an increasing
force during the 
expansion process. Since the driving force $F$ increases with a
diminishing rate when approaching the limit volume change $\Delta
v_{max}$, it follows that the pressure loading case is analogous to
the stress controlled quasistatic loading of bulk specimens. 
According to the simulations, under pressure loading there exists a
critical pressure $P_c$ below which the expansion always stops at a
finite volume and the shell only suffers partial failure  {\it
  (damage)} in the form of cracks but it keeps its integrity. When the
pressure exceeds $P_c$, 
however, the system surpasses the critical volume change $\Delta v_c$
when abruptly a large amount of springs break resulting in the
break-up of the system {\it (fragmentation)}. Note that $P_c << P^{*}_o$.  

The critical volume change $\Delta v_c$ where
fragmentation sets in during the expansion can be identified by the
location of the sudden drop of the elastic energy in Fig.\
\ref{fig:energy_both} caused by the large 
amount of spring breaking which occurs in a very narrow $\Delta v$
interval, resulting in a rapid formation of cracks on the surface. 
The value of $\Delta v_c$ is mainly determined by the fixed breaking
threshold 
$\varepsilon_c$ and the disordered spring properties. 
Since the shell is under constant pressure the nucleated 
microcracks can grow and join giving rise to planar pieces surrounded
by a free crack surface (fragment), as is illustrated in Fig.\
\ref{fig:cracks}$c,d,e)$. First large fragments are formed 
which then break-up into smaller pieces until the surviving springs
can sustain the remaining stress, see Fig.\ \ref{fig:cracks}$c,d,e)$. For
simplicity, in the simulations the pressure is 
kept constant even if the system has lost its integrity, which has
formally the consequence that pieces of the shell formed in the final
state of fragmentation 
keep accelerating under the action of a constant force which explains the
increasing kinetic energy $E_{kin}$ in Fig.\ \ref{fig:energy_both}
following fragmentation. The volume of the system is numerically
calculated as the sum of the volume of pyramidal objects
defined by the surface elements and the center of the sphere, which
provides a meaningful result even after break-up
in Fig.\ \ref{fig:energy_both} in the vicinity of $\Delta v_c$.
The critical pressure $P_c$,
required to exceed the critical volume change $\Delta v_c$ to achieve
fragmentation, can be estimated as $P_c = E_{el}(\Delta
v_c)/(V_o\Delta v_c)$. 

When loading is imposed by an instantaneous energy input $E_o$, 
there is no further energy supply, the
total energy of the system is either constant or decreases due to the
viscous dissipation and the breaking of springs (see Fig.\
\ref{fig:energy_both}). Since the elastic energy $E_{el}$ is solely
determined by the deformation, the curve of $E_{el}$ and the critical
volume change $\Delta v_c$ where break-up arises in Fig.\
\ref{fig:energy_both} coincide with the corresponding values 
of the pressure loading. 
Similarly to the pressure loading case, simulations revealed that a
critical value of the imparted energy $E_c$ can be identified below which
the shell maintains its integrity suffering only damage, while
exceeding $E_c$ gives rise to a complete fragmentation
of the shell.
The resulted fragments on the shell surface obtained in the fragmented
regime can be seen in Fig.\
\ref{fig:cracks}$a,b)$. 

\section{Fragment masses}
To give a quantitative characterization of the break-up of shells and
to reveal the nature of the transition between the damaged and
fragmented states large scale
simulations have been performed varying the control parameters, 
{\it i.e.} the fixed pressure $P_o$, and the imparted energy $E_o$
over a broad range. The most important characteristics of our
fragmenting shell system, that can be compared to the experimental
findings is the variation of fragment masses when changing the control
parameters. 
In the simulations two cut-offs arise for the fragment masses, where
the lower one is defined by the single unbreakable material elements
of the model and the upper one is due to the finite size of the
system. 

For both types of loading above the critical point the typical
fragment size obtained at the instant of break-up decreases with
increasing control parameter, which can 
be described analytically in terms of an energy balance argument similarly
to the one given in Ref.\ \cite{holian1}. 
The loading energy of a shell region of linear extension $L$ and mass $m
\sim L^2$, {\it i.e.} the energy stored in the motion of particles
separating the piece from its surrounding, can be written as
$\left[m/M_{tot}\right]E_{kin}(\Delta v_c)L^2 = \left[E_{kin}(\Delta v_c)/M_{tot}\right]L^4$,
where $E_{kin}(\Delta v_c)$
denotes the total kinetic energy of the shell at the instant of
break-up and $M_{tot}$ is the total mass of the shell. The
separation of the  
piece from its surrounding costs energy proportional to the fragment
surface $\sim L$. 
The equilibrium fragment size can be obtained by
minimizing the sum of the loading and surface energy densities $\rho_E$
\begin{eqnarray}
\label{eq:dens}
\rho_E \sim \frac{E_{kin}(\Delta v_c)}{M_{tot}}L^2 + \frac{1}{L},
\end{eqnarray}
with respect to $L$, which results in $L \sim E_{kin}^{-1/3}$. It has
been shown in the previous section that 
at the critical point $P_c$, $E_c$ the total kinetic energy of the
system when break-up occurs takes zero value $E_{kin}(\Delta v_c) = 0$. It 
follows that above the critical point $E_{kin}$ has a
linear dependence on the distance 
from the critical point so that $E_{kin}(\Delta v_c) \sim (P_o-P_c)$ for
$P_o>P_c$, and $E_{kin}(\Delta v_c) \sim (E_o-E_c)$ for $E_o > E_c$ hold. 
Substituting these
results into Eq.\ (\ref{eq:dens}), the typical fragment mass at the
instant of break-up can be cast into the form
\begin{eqnarray}
\label{eq:typic_pres}
m &\sim& (P_o-P_c)^{-2/3} \ \ \ \mbox{for} \ \ \ P_o>P_c, \\
m &\sim& (E_o-E_c)^{-2/3} \ \ \ \mbox{for} \ \ \ E_o>E_c.
\label{eq:typic_imp}
\end{eqnarray}
\begin{figure}
\begin{center}
\psfrag{aa}{\footnotesize{{\boldmath ${\rm \left<M_{\max}/M_{{\rm tot}}\!\right>}$}}}
\psfrag{bb}{\footnotesize{{\boldmath ${\rm \left<\!M_{\max}^{{\rm 2nd}}/M_{{\rm tot}}\!\right>}$}}}
\epsfig{bbllx=55,bblly=35,bburx=550,bbury=470,file=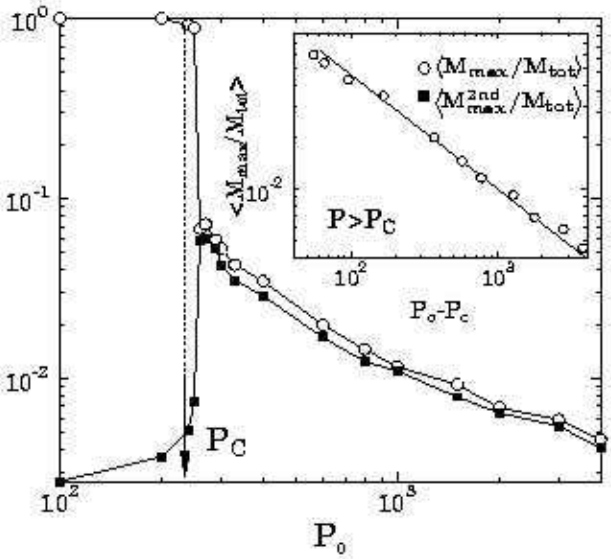,
  width=8.5cm}
 \caption{\small Average mass of the largest
   $\left<M_{max}/M_{tot}\right>$ and second largest
   $\left<M^{2nd}_{max}/M_{tot}\right>$ fragment normalized by 
   the total mass as a function of
   imposed pressure $P_o$. The inset presents a log-log plot of 
   $\left<M_{max}/M_{tot}\right>$ as a 
   function of the distance from the critical point $P_o - P_c$. In
   the main figure the value of $P_c$ is indicated by an arrow.
}
\label{fig:maxmass_press}
\end{center}
\end{figure}
Eqs.\ (\ref{eq:typic_pres},\ref{eq:typic_imp}) express that the
typical fragment mass obtained at the time of
break-up decreases according to a power law with increasing distance
from the critical point. The exponent of the power law is universal in
the sense that it does not depend on specific material properties of
the shell. Later on during 
the fragmentation process the elastic energy stored in deformation
may result in succesive breakings of the large fragments. Hence, it
can be expected that Eqs.\ 
(\ref{eq:typic_pres},\ref{eq:typic_imp}) describe the scaling
behaviour of the largest fragments, which
did not undergo substantial size reduction until reaching the final
relaxed state. 

\begin{figure}
\begin{center}
\psfrag{aa}{\footnotesize{{\boldmath ${\rm \left<M_{\max}\!/\!M_{{\rm tot}}\!\right>}$}}}
\psfrag{bb}{\footnotesize{{\boldmath ${\rm\left<\!M_{\max}^{{\rm 2nd}}/M_{{\rm tot}}\!\right>}$}}}
\epsfig{bbllx=20,bblly=20,bburx=570,bbury=500,file=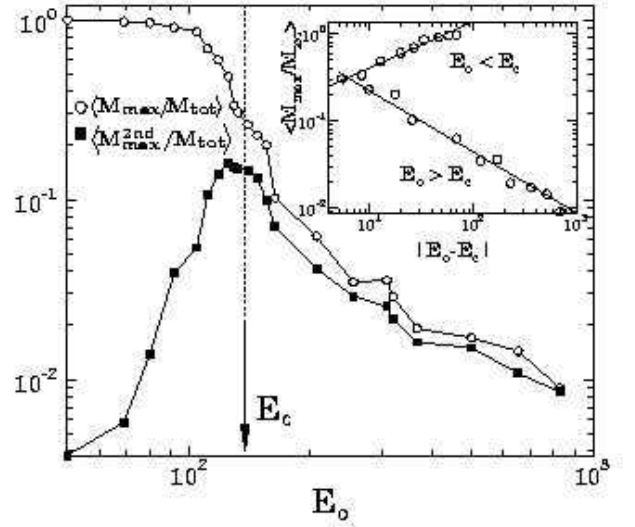,
  width=8.5cm}
 \caption{\small $\left<M_{max}/M_{tot}\right>$ and
   $\left<M^{2nd}_{max}/M_{tot}\right>$ as a function of the
   imparted energy. The inset presents a log-log plot of the largest mass
   as a function 
   of $|E_o-E_c|$, where the increasing and decreasing branches
   characterize the damaged and fragmented states, respectively. The
   location of $E_c$ is indicated in the main figure. 
}
\label{fig:maxmass_inst}
\end{center}
\end{figure}

{\it Largest fragments.} To characterize the degree of
fragmentation, {\it i.e.} the size reduction 
achieved in the simulations, we calculated the average mass of the
largest $\left<M_{max}/M_{tot}\right>$ and of the second largest
$\left<M^{2nd}_{max}/M_{tot}\right>$ fragment normalized by the total mass
as a function of the pressure $P_o$,
and input energy $E_o$ in the case of pressure and impact loading,
respectively \cite{kun1,bhupal}. The results are presented in Figs.\
\ref{fig:maxmass_press}, and \ref{fig:maxmass_inst}.  
It can be seen that in both cases the maximum fragment mass is a
monotonically decreasing function of the control parameters $P_o$ and
$E_o$, however, the functional forms are different in the two cases. 
Low pressure values in Fig.\ \ref{fig:maxmass_press} result in a
breaking of springs, however, hardly any fragments are formed except
for single elements broken out of the shell along cracks. Hence, the
mass of the largest fragment is practically equal to the total mass
$M_{tot}$ of the system, while the second largest fragment is orders
of magnitude smaller ({\it damage}). Increasing however the pressure above the
threshold value $P_c$ the largest fragment mass becomes much smaller
than the total mass, furthermore, in this regime there is only a
slight difference 
between the largest and second largest fragments, indicating the
complete disintegration of the shell into pieces ({\it fragmentation}). 
The value of the critical pressure $P_c$ needed to achieve
fragmentation and the functional form of the curve of
$\left<M_{max}/M_{tot}\right>$ above $P_c$ was 
determined such that $\left<M_{max}/M_{tot}\right>$ was plotted as a
function of the difference $|P_o-P_c|$ varying $P_c$ 
until a straight line is obtained on a double logarithmic plot. 
The result is presented in the inset of Fig.\ \ref{fig:maxmass_press}
where a power law dependence of  $\left<M_{max}/M_{tot}\right>$ is
evidenced as a function of the distance from the critical point
\begin{eqnarray}
  \label{eq:alpha_press}
  \left<M_{max}/M_{tot}\right> \sim |P_o - P_c|^{-\alpha}, \quad
  \mbox{for} \quad P_o > P_c.
\end{eqnarray}
The exponent $\alpha = 0.66 \pm 0.02$ was obtained in
good agreement with the analytic prediction of Eq.\ (\ref{eq:typic_pres}). 
Detailed studies in the vicinity of $P_c$ revealed a finite jump of
both $\left<M_{max}/M_{tot}\right>$ and $\left<M^{2nd}_{max}/M_{tot}\right>$
at $P_c$ which implies that fragmentation
occurs as an abrupt transition at the critical point, see Fig.\
\ref{fig:maxmass_press}. 

\begin{figure}
\begin{center}
\epsfig{bbllx=0,bblly=0,bburx=570,bbury=470,file=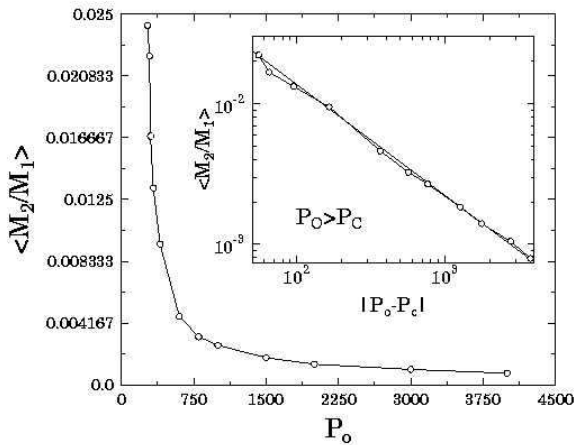,
  width=8.0cm}
 \caption{\small The average fragment mass as a function of the
   imposed pressure $P_o$. 
   The inset presents a log-log plot of the average mass as
   a function of the distance from the critical point $P_o-P_c$ for
   pressure values $P_o > P_c$. The value of $P_c$ is the same as in
   Fig.\ \ref{fig:maxmass_press}.
}
\label{fig:m2m1_pressure}
\end{center}
\end{figure}
In Fig.\ \ref{fig:maxmass_inst} the corresponding results are
presented for the case of impact loading as a function of the total
energy  $E_o$ imparted to the system initially. The mass of the largest
fragment is again a monotonically decreasing function of the control
parameter, however, it is continuous
in the entire energy range considered. Careful
analyzes revealed  the existence of two regimes with a continuous
transition at a critical value of the imparted energy $E_c$. 
In the inset of 
Fig.\ \ref{fig:maxmass_inst}  $\left<M_{max}/M_{tot}\right>$ is shown 
as a function of the distance from the critical point $|E_o -
E_c|$ where $E_c$ was determined using the same technique as
for $P_c$. Contrary  
to the pressure loading, $\left<M_{max}/M_{tot}\right>$ exhibits a
power law behavior on both sides of the critical point but with
different exponents
\begin{eqnarray}
  \label{eq:alpha_inst}
   \left<M_{max}/M_{tot}\right> &\sim& |E_o - E_c|^{\beta}, \quad
  \mbox{for} \quad E_o < E_c, \\
    \left<M_{max}/M_{tot}\right> &\sim& |E_o - E_c|^{-\alpha}, \
  \mbox{for} \quad E_o > E_c.
\end{eqnarray}
The numerical values of the exponents were obtained as $\alpha = 0.66
\pm 0.02$ and $\beta = 0.5 \pm 0.02$, above and below the critical
point respectively. Note that the value of $\alpha$ coincides with the
corresponding exponent of the pressure loading and is in a good
agreement with the analytic prediction of Eq.\
(\ref{eq:typic_imp}). Below the critical point the second largest
fragment is again orders of magnitude smaller than the largest one, 
which implies that in this energy range the shell suffers only damage
in the form of cracks, while above the critical point the break-up of
the entire shell results in comparable values of the largest and second
largest fragment masses. At the transition point $E_c$ between the
damaged and fragmented states the mass of the second largest fragment
has a maximum, while the curve of the largest one exhibits a curvature
change, see Fig.\ \ref{fig:maxmass_inst}.

\begin{figure}
\begin{center}
\epsfig{bbllx=20,bblly=20,bburx=570,bbury=460,file=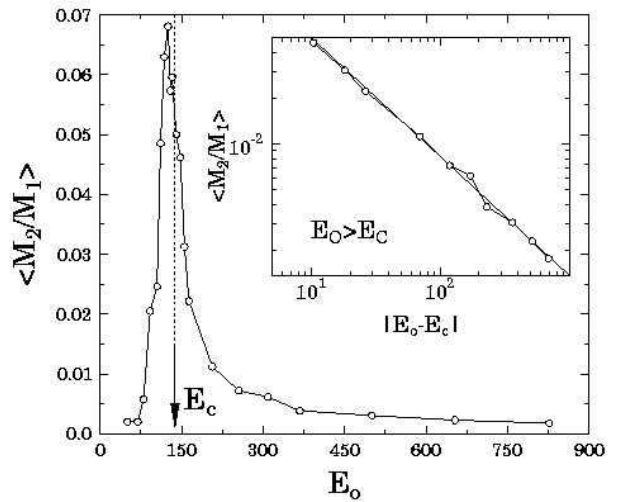,
  width=8.5cm}
 \caption{\small The average fragment mass $\left< M_2/M_1 \right>$ as
   a function of the imparted energy $E_o$. The two regimes can be
   clearly distinguished. The location of the critical point proved to
   be exactly the same as in Fig.\ \ref{fig:maxmass_inst}. The inset
   shows a log-log plot of the average mass as a function of $|E_o -E_c|$.
}
\label{fig:m2m1_inst}
\end{center}
\end{figure}
{\it Average fragment mass.} More insight can be obtained into the
fragmentation process by 
studying the so-called single-event moments of fragment masses 
\begin{eqnarray}
  \label{eq:moments}
  M_k^j = \sum_m m^k n^j(m)-M_{max}^k,
\end{eqnarray}
where $M_k^j$ denotes the $k$th moment of fragment masses $m$ in the
$j$th realization of a fragmentation process, 
$n^j(m)$ is the number of fragments of mass $m$ in event
$j$. The ratio of the second $M_2^j$ and the first
$M_1^j$ moments provides a measure for the average fragment mass in a
specific experiment $j$
\begin{eqnarray}
  \label{eq:avermass}
  \overline{M}^j = \frac{M_2^j}{M_1^j}.
\end{eqnarray}
Averaging over simulations with different realizations of
disorder the average fragment mass $\overline{M} = \left<
  M_2^j/M_1^j\right>$ was obtained as a function of the
control parameter of the system. 

Due to the abrupt nature of the transition from the damaged to the
fragmented states at the critical pressure, under pressure loading 
$\overline{M}$ cannot be evaluated below $P_c$. 
However, when $P_o$ exceeds the critical pressure $P_c$ the
average fragment mass monotonically decreases in Fig.\
\ref{fig:m2m1_pressure}. The inset of Fig.\ \ref{fig:m2m1_pressure} shows 
$\overline{M}$ as a function of the distance from the critical point
$|P_o-P_c|$ where the same value of $P_c$ was used as in Fig.\
\ref{fig:maxmass_press}. A power law dependence of  $\overline{M}$ is
evidenced as a function of $|P_o-P_c|$ 
\begin{eqnarray}
  \label{eq:avermass_pow}
  \overline{M} \sim |P_o-P_c|^{-\gamma},
\end{eqnarray}
for $ P_o > P_c$ and the value of the exponent was obtained to be $\gamma =
0.8\pm0.02$. For impact loading
$\overline{M}$ can be evaluated on both sides of the critical point
with a sharp peak in the vicinity of $E_c$ which 
is typical for continuous phase transitions in finite systems, see Fig.\
\ref{fig:m2m1_inst}. A power law 
dependence of $\overline{M}$ on the distance from the critical point
\begin{eqnarray}
\label{eq:avermass_inst}
\overline{M} \sim |E_o - E_c|^{-\gamma}
\end{eqnarray}
is again revealed for $E_o > E_c$, which is illustrated in the 
inset of Fig.\ \ref{fig:m2m1_inst}. The value of the exponent was
determined by fitting $\gamma = 0.79\pm 0.02$, which practically
coincides with the $\gamma$ value of pressure loading. 

\begin{figure}
\begin{center}
\epsfig{bbllx=20,bblly=20,bburx=570,bbury=500,file=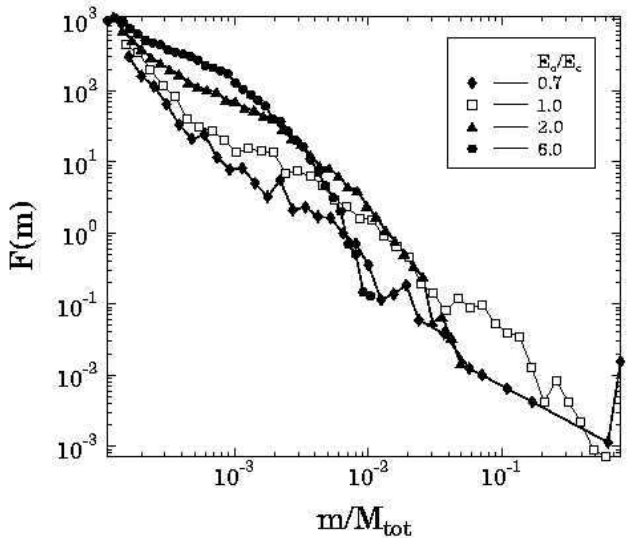,
  width=8.5cm}
 \caption{\small Mass distribution of fragments at various energies
   below and above the critical point. 
}
\label{fig:massdist_inst}
\end{center}
\end{figure}
{\it Fragment mass distributions.} The most important characteristic
quantity of our system which can 
also be compared to the experimental results is the mass distribution
of fragments $F(m)$.
Under impact loading for $E_o < E_c$ we found that $F(m)$
has a pronounced peak at large fragments indicating the presence of
large damaged pieces, see Fig.\ \ref{fig:massdist_inst}. Approaching
the critical point $E_c$ the peak gradually disappears and the
distribution asymptotically becomes a power law at $E_c$. We can
observe in Fig.\ \ref{fig:massdist_inst} that above the
critical point the power law remains for small fragments followed by
a cut-off for the large ones, which decreases with increasing $E_o$. 
 
\begin{figure}
\begin{center}
\epsfig{bbllx=20,bblly=20,bburx=570,bbury=480,file=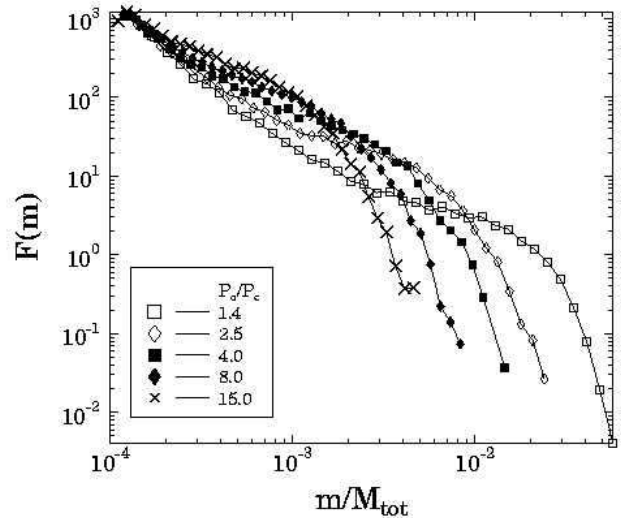,
  width=8.5cm}
 \caption{\small Mass distribution of fragments at various pressure values.
}
\label{fig:massdist_press}
\end{center}
\end{figure}

For pressure loading $F(m)$ can only be evaluated above $P_c$. The
evolution of $F(m)$ with increasing pressure is presented in Fig.\
\ref{fig:massdist_press}, where the mass
distribution always shows a power law behavior for small fragments
with a relatively broad cut-off for the large ones.
For the purpose of comparison, a mass
distribution $F(m)$ obtained at an impact energy close to the critical
point $E_c$, and distributions at two different pressure values $P_o$
of the ratio 1.6 are plotted in Fig.\ \ref{fig:exp_imp} along with the
experimental results. For impact an excellent agreement with the
experimental and theoretical results is evidenced. For pressure
loading, the functional form of $F(m)$ has a nice qualitative
agreement with the experimental findings on the explosion of eggs,
furthermore, distributions at the same ratio of pressure
values obtained by simulations and experiments show the
same tendency of evolution, see Fig.\ \ref{fig:exp_imp}.

\begin{figure}
\begin{center}
\psfrag{aa}{{\large {\boldmath ${\rm m/\overline{M}}$}}}
\psfrag{bb}{{\large {\boldmath ${\rm  F(m)\cdot \overline{M}^{\delta}}$}}}
\epsfig{bbllx=20,bblly=20,bburx=560,bbury=480,file=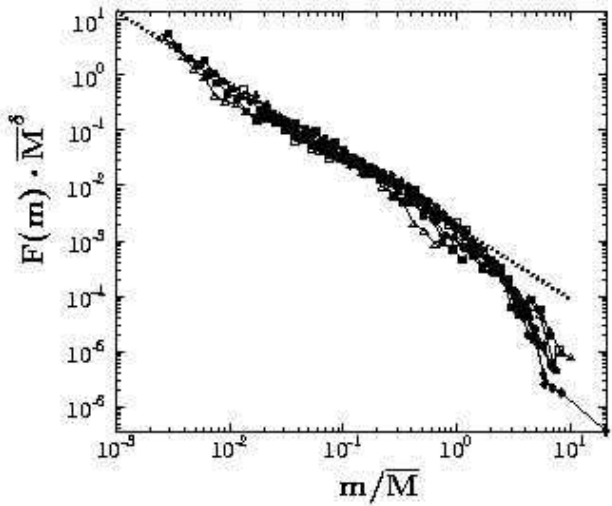,
  width=8.5cm}
 \caption{\small Rescaled plot of the mass distributions for imparted
   energies above the critical point $E_o >E_c$. The dashed line
   shows the fitted power law with an exponent $\tau = 1.35 \pm 0.03$.
}
\label{fig:mass_scal_inst}
\end{center}
\end{figure}

\begin{figure}
\begin{center}
\psfrag{aa}{{\large {\boldmath ${\rm m/\overline{M}}$}}}
\psfrag{bb}{{\large {\boldmath ${\rm F(m)\cdot \overline{M}^{\delta}}$}}}
\epsfig{bbllx=20,bblly=20,bburx=560,bbury=490,file=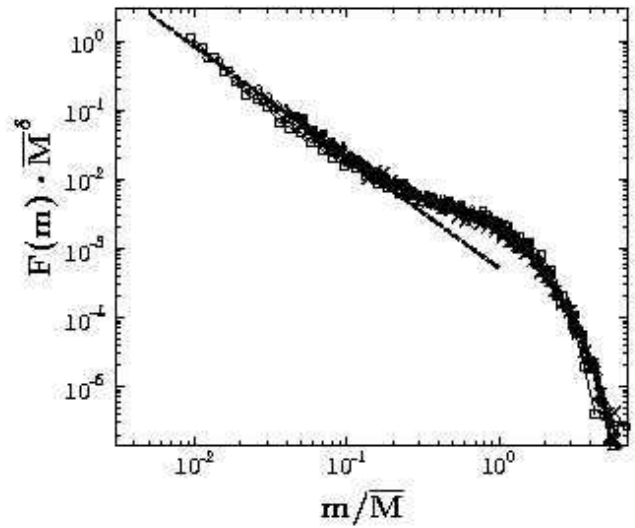,
  width=8.8cm}
 \caption{\small Rescaled plot of the mass distributions of various
   pressure values above the critical point $P_o > P_c$. The dashed
   line indicates the fitted power law with an exponent $\tau = 1.55
   \pm 0.03$.
}
\label{fig:mass_scal_press}
\end{center}
\end{figure}

Figs.\  \ref{fig:mass_scal_inst} and \ref{fig:mass_scal_press} demonstrate
that by 
rescaling the mass distributions above the critical point by plotting
$F(m) \cdot \overline{M}^{\delta} $ as a function of $m/\overline{M}$ an
excellent data collapse is obtained with $\delta = 1.6\pm 0.03$. The
data collapse implies the validity of the scaling form 
\begin{eqnarray}
\label{eq:scaling}
F(m) \sim m^{-\tau} \cdot f(m/\overline{M}), 
\end{eqnarray}
typical for critical
phenomena. The cut-off function $f$ has a simple exponential form
$\exp{\left(-m/\overline{M}\right)}$ for impact loading (see Fig.\
\ref{fig:mass_scal_inst}), and a more complex one 
containing also an exponential component for the pressure case (see
Fig.\ \ref{fig:mass_scal_press}).
The average fragment mass $\overline{M}$ occurring in the scaling form
Eq.\ (\ref{eq:scaling}) diverges according to a power law given by Eqs.\
(\ref{eq:avermass_pow},\ref{eq:avermass_inst}) when 
approaching the critical point. The good quality of collapse and the
functional form Eq.\ (\ref{eq:scaling}) also imply that the exponent
$\tau$ of the mass distribution  does not depend on the value of the
pressure $P_o$ or the kinetic energy $E_o$ contrary to the bulk
fragmentation where an energy dependence of $\tau$ was reported
\cite{ching}. 

The rescaled plots make possible an accurate determination of the
exponent $\tau$, where $\tau = 1.35\pm 0.03$ and $\tau = 1.55\pm 0.03$
were obtained for impact and pressure loading, respectively.
Hence, a good quantitative agreement of the theoretical and experimental
values of the exponent $\tau$ is
evidenced for the impact loading of shells, however, for the case of
pressure loading the numerically obtained exponent turned out to be
somewhat higher than in the case of exploded eggs.

\section{Discussion and outlook}
We presented a detailed experimental and theoretical study of the
break-up of closed shells arising due to a shock inside the shell. 
For the purpose of experiments brown and white hen egg-shells were
carefully prepared to ensure a high degree of brittleness of the
disordered shell material. The break-up of the shell was studied under
two different loading conditions, {\it i.e.} explosion caused by
a combustible mixture and impact with the hard ground.
As the main outcome of the experiments, the mass
distribution of fragments proved to be a power law in both loading
cases for small fragment sizes, however, qualitative differences were
obtained in the limit of large fragments for the shape of the cut-off. 

We worked out a discrete element model for the break-up of shells 
which provides an insight into the dynamics of the process by
simultaneously monitoring several microscopic quantities in the
framework of molecular dynamics simulations. 
In the simulations two ways of loading have been considered, which
mimic the experimental conditions and represent limiting cases of
energy input rates: during an expansion under constant pressure $P_o$
the shell is driven by an increasing force 
with a continuous increase of the imparted energy, while the impact 
loading realizes the instantaneous input of an energy
$E_o$. 
Simulations revealed that 
depending on the value of $P_o$ and $E_o$, the final outcome
of the break-up process can be classified into two states, {\it i.e.}
damaged and fragmented with a sharp transition in between at a
critical value of the control parameters $P_c$ and $E_c$.
In the fragmented regime power law fragment mass distributions were
obtained in satisfactory agreement with the experimental 
findings. 
Analyzing the behavior of the system in the vicinity of the
critical point $P_c$, $E_c$, we showed that power law distributions
arise in the break-up of shells due to an underlying phase transition
between the damaged and fragmented states, which proved to be abrupt
for explosion, and continuous for impact. 

Due to its unique characteristics, the break-up of shells defines a
new universality class of fragmentation phenomena, different from that
of the two- and three-dimensional bulk systems. Based on
universality, our results should be applicable to describe
the break-up of other closed shell systems composed of disordered brittle
materials. Explosion of shell-like fuel containers, tanks, high
pressure vessels often occur as accidental events in industry, or in
space missions where also the explosion of complete satellites may
occur creating a high amount of space debris orbiting about Earth. For
the safety design of shell constructions, and for the tracking of
space debris it is crucial to have a comprehensive understanding of
the break-up of shells. Due to the universality of fragmentation
phenomena, our results can be exploited for these purposes.

In the fragmentation of bulk systems under appropriate conditions
a so-called detachment effect is observed when a
surface layer breaks off from the bulk and undergoes a separated
fragmentation process \cite{ferenc1,ching}. This
effect also shows up
in the fragment mass distributions in the form of a power law regime
of small fragments of an exponent smaller than for the large ones.
Our results on shell fragmentation can also provide a
possible explanation of this kind of composite power laws of bulk
fragmentation \cite{ferenc1,ching}.

\begin{acknowledgments}
This work was supported by the Collaborative Research Center SFB381
and by OTKA T037212, M041537. F.\ Kun
was supported by the Research Contract FKFP 
0118/2001 and by the Gy\"orgy B\'ek\'esi Foundation of the Hungarian
Academy of Sciences. The authors are also thankful to the technical
support of H.\ Gerhard from IKP. 
\end{acknowledgments}

\end{document}